\documentclass[aps,prl,twocolumn,superscriptaddress,showpacs]{revtex4}
\usepackage{graphicx}
\begin{document}

%\preprint{}

{\bf Comment on ``Quantum Opacity, the RHIC HBT Puzzle, and the Chiral
Phase Transition''}
\vspace{1mm}

%\author{\input{Phobos_authors_current}}
%\thanks{}
%\affiliation{}


%\date{\today}

%\begin{abstract}
In a recent Letter, Cramer, Miller, Wu and Yoon \cite{cramer} (CMWY) presented 
a model based on a relativistic quantum mechanical treatment of opacity
and refractive effects on momentum correlations between pairs of identical
particles and particle transverse momentum ($p_T$) spectra. The values
of the ten model parameters were obtained from a simultaneous fit to 
the HBT radii and pion $p_T$ spectra measured for central Au+Au collisions at
$\sqrt{s_{_{NN}}} =$ 200~GeV. CMWY conclude from the fitted values of their
model parameters that the data are consistent with the emission of pions 
from a 
dense medium in which chiral symmetry has been restored. In addition,
the authors provide predictions of the momentum dependence
of the HBT radii and pion $p_T$ spectra in the region below
200 MeV/c. However, the authors chose not to compare their model
predictions with the published
experimental data for pion yields at very low $p_T$ obtained using the 
PHOBOS detector at RHIC. The purpose of this
Comment is to make the comparison of the CMWY model predictions 
with the PHOBOS data.

\begin{figure}
\includegraphics[width=7.2cm]{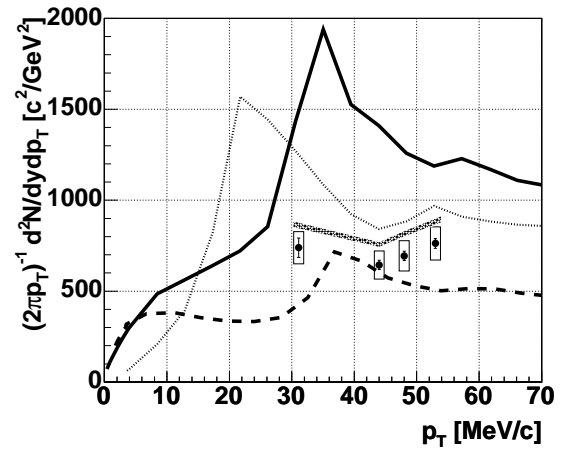}
\caption{\label{fig:fig1} $p_T$ spectra of charged pions. Points 
show PHOBOS published data for ($\pi^{+}+\pi^{-}$)/2  yields measured in the 
15\%
most central Au+Au collisions at $\sqrt{s_{_{NN}}} =$ 200~GeV. 
Bars denote statistical 
errors and the boxes correspond to  90\% C.L. systematic uncertainties. 
The crossed curve shows the extent to which the yield is expected to increase
(15-20\%) for a more central event selection in the data. 
Other curves show results of the CMWY
model calculations with the dashed curve representing the published results
\cite {cramer}. The dotted and solid curves  represent
more recent model predictions \cite{cramer_private}. See text 
for more details.}
\end{figure}

The PHOBOS experiment has a unique capability of measuring charged particles
down to very low $p_T$. The results \cite{phostop} reach $p_T$ as
low as 30 MeV/c for charged pions, and cover the region where the CMWY
model predicted notable structure in the pion $p_T$ spectrum. The 
published model predictions are shown as a dashed curve in Fig.~\ref{fig:fig1}.
However, in the 
original paper \cite{cramer} there was an error in the representation
of the STAR pion momentum spectra \cite{star}, consequently leading to the
extraction of incorrect values of the 
model parameters. As a result, the dashed curve, shown in Fig.~\ref{fig:fig1},
disagrees with the STAR data at higher $p_T$.
This mistake was pointed out to CMWY by the authors of this comment and, 
as a result,
a new set of model predictions for the spectral behavior
at very low transverse momenta was generated \cite{cramer_private}. 
The corrected model predictions, shown by a dotted  curve in
Fig.~\ref{fig:fig1}, differ significantly from those published in 
\cite{cramer}. More
recently, an alternative fit was performed in which the pion momentum
distribution used in the fit was modified ``to remove from the spectrum
those pions that did not participate in the Bose-Einstein symmetrization''
\cite{cramer_private}. The results from this version of the fit are shown 
as the solid curve in Fig.~\ref{fig:fig1}.
It is clear that the predicted yield at low $p_T$ is very sensitive to the 
values of the parameters and the details of the model.

The low-transverse momentum charged pion yields, measured by PHOBOS 
\cite{phostop}, are overlayed on the model curves in Fig.~\ref{fig:fig1}.
Since our analysis of
particle production in this momentum region does not allow for 
separation of positive and negative pions, the invariant
$p_T$ yields measured for charged pions are divided
by 2. It should be noted that our results correspond to the
15\% most central Au+Au collisions. The centrality of the predictions
is not clearly specified, but we infer from the data shown that they are
for the 5\% most central collisions. The number of pions per event is expected
to be higher for more central data. Thus, imposing a more stringent
centrality selection leads to yields higher than
those measured by PHOBOS. One can roughly estimate the size of this increase 
from the $p_T$ spectra measured by PHENIX at higher $p_T$ for centrality bins
of 0-5\% and 0-15\% \cite{phenix} or from the centrality dependence of the 
$p_T$ integrated charged particle $dN/d\eta$ yields measured by PHOBOS 
\cite{phobosmult}. Both
estimates give scaling factors in a range from 15 to 20\%. 
Interpolation of the STAR
published data at higher $p_T$ \cite{star} for centrality bins of 0-10\% 
and 0-20\% suggests a similar
increase.
Multiplying our measured points
by factors of 1.15 to 1.20 gives the thick crossed curve shown in 
Fig.~\ref{fig:fig1}.

It is clear that it would be best to test models, such as those 
discussed here, in
a range of centralities. This requires more  central data, which will be
provided by PHOBOS, as well as model calculations for less central 
collisions.
%One can see that the most recent model
%predictions(solid curve) exceed the data by more than 30\% even after 
%accounting for differences in centrality. 
We conclude that the low $p_T$ pion yields uniquely measured
by PHOBOS provide valuable information for evaluating models of 
the dense medium created in heavy ion collisions
at RHIC energies.

PHOBOS Collaboration: \\  
B.B.Back$^1$,
M.D.Baker$^2$,
M.Ballintijn$^4$,
D.S.Barton$^2$,
R.R.Betts$^6$,
A.A.Bickley$^7$,
R.Bindel$^7$,
A.Budzanowski$^3$,
W.Busza$^4$,
A.Carroll$^2$,
M.P.Decowski$^4$,
E.Garc\'{\i}a$^6$,
N.George$^{1,2}$,
K.Gulbrandsen$^4$,
S.Gushue$^2$,
C.Halliwell$^6$,
J.Hamblen$^8$,
G.A.Heintzelman$^2$,
C.Henderson$^4$,
D.J.Hofman$^6$,
R.S.Hollis$^6$,
R.Ho\l y\'{n}ski$^3$,
B.Holzman$^2$,
A.Iordanova$^6$,
E.Johnson$^8$,
J.L.Kane$^4$,
J.Katzy$^{4,6}$,
N.Khan$^8$,
W.Kucewicz$^6$,
P.Kulinich$^4$,
C.M.Kuo$^5$,
W.T.Lin$^5$,
S.Manly$^8$,
D.McLeod$^6$,
A.C.Mignerey$^7$,
R.Nouicer$^6$,
A.Olszewski$^3$,
R.Pak$^2$,
I.C.Park$^8$,
H.Pernegger$^4$,
C.Reed$^4$,
L.P.Remsberg$^2$,
M.Reuter$^6$,
C.Roland$^4$,
G.Roland$^4$,
L.Rosenberg$^4$,
J.Sagerer$^6$,
P.Sarin$^4$,
P.Sawicki$^3$,
W.Skulski$^8$,
S.G.Steadman$^4$,
P.Steinberg$^2$,
G.S.F.Stephans$^4$,
A.Sukhanov$^2$,
J.-L.Tang$^5$,
A.Trzupek$^3$,
C.Vale$^4$,
G.J.van~Nieuwenhuizen$^4$,
R.Verdier$^4$,
F.L.H.Wolfs$^8$,
B.Wosiek$^3$,
K.Wo\'{z}niak$^3$,
A.H.Wuosmaa$^1$,
B.Wys\l ouch$^4$\\
\noindent
{\footnotesize
$^1$~Argonne National Laboratory, Argonne, IL 60439\\
$^2$~Brookhaven National Laboratory, Upton, NY 11973\\
$^3$~Institute of Nuclear Physics PAN, Krak\'{o}w, Poland\\
$^4$~Massachusetts Institute of Technology, Cambridge, MA 02139\\
$^5$~National Central University, Chung-Li, Taiwan\\
$^6$~University of Illinois at Chicago, Chicago, IL 60607\\
$^7$~University of Maryland, College Park, MD 20742\\
$^8$~University of Rochester, Rochester, NY 14627\\}

%\begin{acknowledgments}
%This work was partially supported by U.S. DOE grants DE-AC02-98CH10886,
%DE-FG02-93ER40802, DE-FC02-94ER40818, DE-FG02-94ER40865, DE-FG02-99ER41099, 
%and
%W-31-109-ENG-38 as well as NSF grants 9603486, 9722606 and 0072204.  
%The Polish
%group was partially supported by KBN grant 2-P03B-10323.  The NCU group was
%partially supported by NSC of Taiwan under contract NSC 89-2112-M-008-024.
%\end{acknowledgments}

%\bibliography{stopping_v110503.bib}

\end{document}